\documentstyle[epsf,aps]{revtex}

\begin{document}

\twocolumn[\hsize\textwidth\columnwidth\hsize\csname
@twocolumnfalse\endcsname

\title{Supernova Neutrinos, Neutrino Oscillations, and the Mass of the
Progenitor Star} \author{Keitaro Takahashi$^{\rm a}$, Katsuhiko
Sato$^{\rm a,b},$
        Adam Burrows$^{\rm c}$ and Todd A. Thompson$^{\rm d,e}$\\ {\it
$^{\rm a}$Department of Physics, University of Tokyo, 7-3-1 Hongo,
Bunkyo,}\\ {\it Tokyo 113-0033, Japan}\\ {\it $^{\rm b}$Research Center
for the Early Universe, 
University of Tokyo,}\\
{\it 7-3-1 Hongo, Bunkyo, Tokyo 113-0033, Japan}\\
{\it $^{\rm c}$Steward Observatory, The University of Arizona,}\\ {\it
Tucson, AZ 85721}\\ {\it $^{\rm d}$Astronomy Department and Theoretical
Astrophysics Center, 601 Campbell Hall,}\\ {\it The University of
California, Berkeley, CA 94720}\\ {\it $^{\rm e}$ Hubble Fellow}}

\date{\today}

\maketitle

\begin{abstract}

We investigate the initial progenitor mass dependence of the early-phase
neutrino signal 
from supernovae taking neutrino oscillations into account. The
early-phase 
analysis has advantages in that it is not affected by the time evolution
of the density structure of the star due to shock propagation or whether  
the remnant is a neutron star or a black hole.
The initial mass affects the evolution of the massive star and 
its presupernova structure, which is important for 
two reasons when considering the neutrino signal. 
First, the density profile of the mantle affects the dynamics of 
neutrino oscillation in supernova. Second, the final iron core structure
determines the features of the neutrino burst, i.e., the luminosity and
the average energy. We find that both effects are rather small. This is
desirable when we try to extract information on neutrino parameters from
future supernova-neutrino observations. Although the uncertainty due to
the progenitor mass is not small for intermediate $\theta_{13}$ 
($10^{-5} \lesssim \sin^{2}{2 \theta_{13}} \lesssim 10^{-3}$), we can,
nevertheless, determine the character of the mass hierarchy 
and whether $\theta_{13}$ is very large or very small.

\medskip
\noindent
$PACS:$ 14.60.Pq; 14.60.Lm; 96.40.Tv; 97.60.Bw; 

\noindent
$Keywords:$ Neutrino oscillations; Supernovae;

\end{abstract}

\vskip2pc]

\vskip1cm

\section{Introduction}

Core-collapse supernovae produce a huge flux of neutrinos of all types.
The observation of neutrinos from SN1987A in the Large Magellanic Cloud
was a milestone for neutrino astronomy \cite{K2_SN1987a,IMB_SN1987a}.
Although some important constraints on neutrino properties were obtained
\cite{Arafune,Sato,Goldman,Jegerlehner1996,LunardiniSmirnov,Minakata2001},
the total of 19 events in the Kamiokande and IMB detectors was 
frustratingly sparse. On the other hand, SuperKamiokande is expected to 
detect about five to ten thousand neutrinos from the next galactic
supernova. Together with the other neutrino detectors arrayed around the world
(e.g. SNO, LVD, ICARUS, IceCube), a high-statistics (nearby) supernova 
will provide an enormous amount of information on both supernova and 
neutrino properties.

Despite the new oscillation data from solar \cite{Fukuda2001,SNO}, 
atmospheric \cite{Fukuda1999}, reactor \cite{KamLAND,Apollonio1999}, and
accelerator \cite{K2K} neutrinos, the neutrino mixing angle $\theta_{13}$ 
and the nature of the mass hierarchy remain unknown.  The constraints and 
implications of these data have been explored by many authors 
\cite{Dighe2000,Fogli2002,LunardiniSmirnov2003,DigheKeilRaffelt1,DigheKeilRaffelt2,KT1,KT2,KT3,KT4}. The greatest uncertainty for the study of neutrino 
oscillations in the supernova context is the spectral
and temporal evolution of the neutrino burst itself.   
Although the hierarchy of the average energies of the 
$\nu_{e}, \bar{\nu}_{e}$, and $\nu_{x}$ neutrinos
($\langle E_{\nu_{e}} \rangle <  \langle E_{\bar{\nu}_{e}} \rangle 
< \langle E_{\nu_{x}} \rangle$) is believed to hold, quantitatively
their specific spectra remain a matter of detailed calculation 
\cite{Raffelt2001,Raffelt2003}. ($\nu_{x}$ denotes $\nu_{\mu},\nu_{\tau}
$ and their antineutrinos.) Since a supernova is the end-product of a
massive star, the properties of the neutrino burst will depend on the
properties of the progenitor star, in particular its initial mass and
envelope structure \cite{MayleD,Mayle1987}.

In this paper, we study how the initial mass of the progenitor star
affects the early neutrino burst and the signature of neutrino
oscillations in supernovae. The rest of the paper is organized as
follows. In \S II, we briefly review the evolution of massive stars,
supernova explosions, and neutrino bursts. Conversion probabilities 
for progenitor stars with various masses are calculated and compared 
in \S III. Then, neutrino spectra with and without
neutrino oscillations are shown in \S IV. Finally in \S V we discuss 
the ramificatons of our results and summarize our conclusions.

\section{Evolution of Massive Stars and
Supernovae\label{section:evolution}}

Here, we briefly summarize the evolution of massive stars and supernovae. 
See \cite{WoosleyHegerWeaver2002} for a more extended review.

The evolution and death of a single massive star are determined by its 
initial mass and metallicity. For the metallicity of the Sun, a star
with initial mass below about 8 $M_{\odot}$ does not ignite carbon
burning and forms a white dwarf. Above about 8 $M_{\odot}$, 
a star completes the advanced burning stages, including for most stars
silicon burning. This leads to the presupernova state, which is
characterized by a core (iron or NeOMg) of roughly the Chandrasekhar
mass surrounded by active burning shells and the accumulated ashes of
oxygen, neon, carbon, and/or helium burning. The degenerate core in
excess of the Chandrasekhar mass will dynamically collapse and explode 
as a supernova.

Up to an initial mass of $20 - 25 M_{\odot}$, the supernova leaves
behind a neutron star \cite{Fryer1999,FryerKalogera2001}.
Almost all of the binding energy of the neutron star, \begin{equation}
E_{\rm b} \simeq \frac{G M_{\rm NS}^{2}}{2R_{\rm NS}} \simeq 1.5\times
10^{53} {\rm erg} 
\left( \frac{M_{\rm NS}}{M_{\odot}} \right)^{2}
\left( \frac{10 {\rm km}}{R_{\rm NS}} \right),
\end{equation}
is radiated away as neutrinos. Here $G$, $M_{\rm NS}$ and $R_{\rm NS}$
are the gravitational constant, the protoneutron star's mass, and its radius, 
respectively. Due to the differences in interaction strength, average 
neutrino energies vary with neutrino flavor.  
A full-scale numerical simulation by the Livermore 
group \cite{Wilson1986} using a multi-group flux-limiter diffusion
approach for neutrino transport 
finds $\langle E_{\nu_{e}} \rangle \simeq 13\ {\rm MeV}, \;\; \langle
E_{\bar{\nu}_{e}} \rangle \simeq 16\ {\rm MeV}, \;\; \langle E_{\nu_{x}}
\rangle \simeq 23\ {\rm MeV}$ and almost perfect equipartition of the
luminosities \cite{Totani1998}. Simulations with more sophisticated
neutrino transport \cite{Burrows}, which are, however, limited to the
early phase of neutrino burst, derive a smaller flavor-dependence 
for the average energy. 
These differences are essential if neutrino oscillations are to have a 
perceptible effect on supernova neutrino detection.
(For details on supernova neutrinos, see, for example, the reviews by
Burrows\cite{Burrows2} and Suzuki\cite{Suzuki}.)

For progenitor masses larger than $20 - 25 M_{\odot}$, sufficient mass
may fall back onto the protoneutron star after explosion to turn it into
a black hole. Even in this case, during the early phase which we
consider in this paper, the properties of the neutrino burst are almost 
the same as stated above.

The mass of the progenitor star affects the neutrino oscillation
signature of supernova neutrinos through differences in the mantle and
core structures. The density profile of the progenitor star, especially 
of the mantle, is important because it is related to the dynamics of 
neutrino flavor conversion. On the other hand, the structure of 
the iron core at the collapse determines the characteristics of 
the neutrino burst, e.g., the average energy and luminosity for each flavor. 
We discuss these in the following sections.

\section{Neutrino Flavor Conversion and the Density Profile
\label{section:conversion}}

\begin{figure}
\begin{center}
\epsfxsize=6.5cm
\epsffile{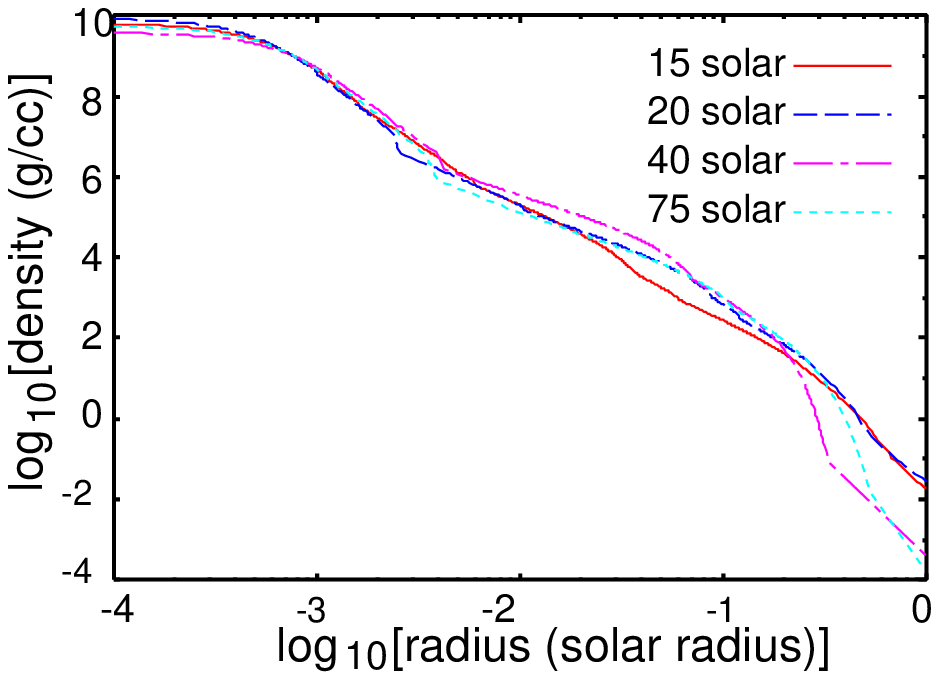}
\end{center}
\caption{Density profiles of stars just before supernova explosion with
initial mass $15 M_{\odot}, 20 M_{\odot}, 40 M_{\odot}$ and $75
M_{\odot}$. \label{fig:density}} \end{figure}

The evolution of a massive star is significantly affected by mass loss
due to a stellar wind. Indeed, the mass loss can become so strong for a
star with initial mass more than about 35 $M_{\odot}$ and solar
metallicity that the entire hydrogen envelope can be lost prior to 
the explosion of the star. It is suggested in \cite{WoosleyHegerWeaver2002} 
that the maximum in the final mass is about 20 $M_{\odot}$. 
In addition, the density profile of a star
just before the supernova explosion will not depend much on the initial
mass of the star. The final density profiles of stars with various
initial masses are shown in Fig. \ref{fig:density} \cite{SCIhomepage}. 
As is expected, they are similar just before collapse.

The density structure in the range 
$10 {\rm g/cc} \lesssim \rho \lesssim 10^{4} {\rm g/cc}$ 
is particularly important when considering neutrino oscillations. 
Neutrinos produced in the high-density region of the iron
core interact with matter before emerging from the supernova. Due to the
non-zero masses and non-zero vacuum mixing angles  among various
neutrino flavors, flavor conversions can occur in supernovae. When the
mixing angle is small, these conversions occur mainly in the resonance
layer, where the density is \begin{eqnarray} \rho_{\rm res} & \simeq &
1.4 \times 10^{6} {\rm g/cc} \left( \frac{\Delta m^{2}}{1 {\rm eV}^{2}}
\right) \left( \frac{10 {\rm MeV}}{E_{\nu}} \right) \nonumber \\ & &
\times \left( \frac{0.5}{Y_{e}} \right) \cos{2 \theta}. \end{eqnarray}
$\Delta m^{2}$ is the mass squared difference, $\theta$ is the mixing
angle, $E_{\nu}$ is the neutrino energy, and $Y_{e}$ is the mean 
number of electrons per baryon. Since the inner supernova core is too
dense to allow resonance conversion, we focus on two
resonance points in the outer supernova envelope. One that occurs at 
higher density is called the H-resonance and the other, which occurs at
lower density, is called the L-resonance. If the mass hierarchy is
normal, both resonances occur in the neutrino sector. On the other hand,
if the mass hierarchy is inverted, the H-resonance occurs in the
antineutrino sector and the L-resonance occurs in the neutrino sector.

The dynamics of conversion including large mixing case is
determined by the adiabaticity parameter $\gamma$, 
\begin{equation}
\gamma \equiv \frac{\Delta m^{2}}{2 E_{\nu}} 
\frac{\sin^{2}{2 \theta}}{\cos{2 \theta}} 
\frac{n_{e}}{|dn_{e}/dr|},
\label{gamma}
\end{equation}
which depends on the mixing angle and
the mass-squared difference between the involved flavors.
In eq. \ref{gamma}, ${n_{e}}$ is the electron number density. We define
these as: \begin{equation} \theta_{13} \; {\rm and} \; \Delta
m^{2}_{13}, \; {\rm at\ the \; H-resonance}, \end{equation} and 
\begin{equation}
\theta_{12} \; {\rm and} \; \Delta m^{2}_{12}, \; {\rm at\ the \;
L-resonance}. \end{equation} Here, the Cabibbo-Kobayashi-Maskawa (CKM)
matrix is taken as: \begin{eqnarray} && U  =  \nonumber \\ &&
\left(\begin{array}{ccc} c_{12}c_{13} & s_{12}c_{13} & s_{13}\\
-s_{12}c_{23}-c_{12}s_{23}s_{13} & c_{12}c_{23}-s_{12}s_{23}s_{13} 
& s_{23}c_{13}\\
s_{12}s_{23}-c_{12}c_{23}s_{13} & -c_{12}s_{23}-s_{12}c_{23}s_{13} 
& c_{23}c_{13} \nonumber \end{array}\right)\label{mixing_matrix},
\end{eqnarray}
where $s_{ij} = \sin{\theta_{ij}}, c_{ij} = \cos{\theta_{ij}}$ 
for $i,j=1,2,3\, (i<j)$.
When $\gamma \gg 1$, the resonance is referred to as an `adiabatic
resonance' and 
the fluxes of the two involved mass eigenstates are completely
exchanged. 
However, when $\gamma \ll 1$, the resonance is called `nonadiabatic' and
the conversion does not occur. 
The dynamics of the resonance in supernovae is surveyed by Dighe and
Smirnov \cite{Dighe2000}.

\begin{figure}
\begin{center}
\epsfxsize=6.5cm
\epsffile{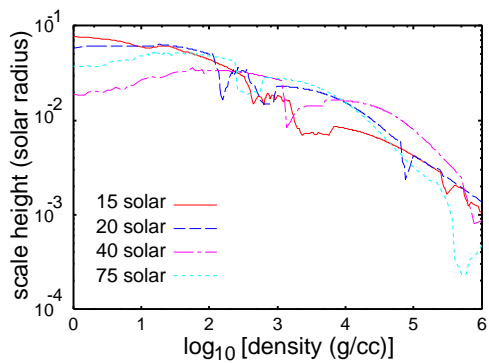}
\end{center}
\caption{Scale heights $n_{e}/|dn_{e}/dr|$ of stars just before the
supernova 
explosion with initial masses $15\ M_{\odot}, 20\ M_{\odot}, 40\
M_{\odot}$, and $75\ M_{\odot}$. \label{fig:gradient}} \end{figure}

The density profile of the star comes into the adiabaticity parameter 
$\gamma$ as the scale height $n_{e}/|dn_{e}/dr|$.  A smaller scale
height, that is, a steeper density profile results in less adiabatic
resonance. The scale heights of stars with various initial masses,
calculated from the density profiles shown in Fig. \ref{fig:density},
are given in Fig. \ref{fig:gradient} as functions of the density.
The scale heights vary significantly, but differences between various
initial masses are factors of 2 or 3 at the densities relevant for
resonances.

To estimate the effects of these differences on neutrino oscillations,
we solve numerically the evolution equations for the neutrino wave
functions along the density profiles shown in Fig. \ref{fig:density}. We
take the following values for the neutrino parameters: \begin{eqnarray}
\sin^{2}{2 \theta_{12}} = 0.84, & \;\;\;\; & 
\Delta m^{2}_{12} = 7 \times 10^{-5} {\rm eV}^{2}, \nonumber \\ \sin^{2}
{2 \theta_{23}} = 1.0, & \;\;\;\; & 
\Delta m^{2}_{23} = 3.2 \times 10^{-3} {\rm eV}^{2}. \end{eqnarray} The
values of $\theta_{12}$ and $\Delta m^{2}_{12}$ are taken from the
global analysis of the solar neutrino observations and the KamLAND
experiment\cite{Bahcall2003} and correspond to the large mixing angle
(LMA) solution of the solar neutrino problem, while those of $\theta_{23}$ and 
$\Delta m^{2}_{23}$ are taken from the analysis of the atmospheric
neutrino observations\cite{Fukuda1999}. As for $\theta_{13}$, we take
$\sin^{2}{2 \theta_{13}} = 10^{-4}$. With this $\theta_{13}$, the
H-resonance is neither perfectly adiabatic nor nonadiabatic and its
adiabaticity is sensitive to the density profile. 
The normal hierarchy is assumed, but the results are the same for the
inverted hierarchy if $\nu_{e}$ is replaced by $\bar{\nu}_{e}$.

\begin{figure}
\begin{center}
\epsfxsize=6.5cm
\epsffile{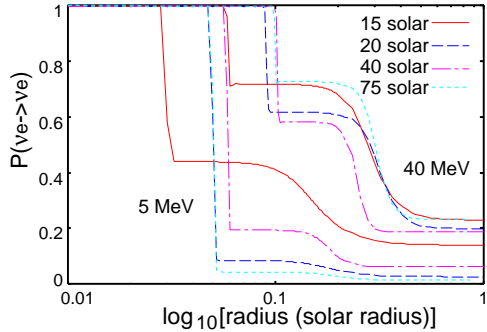}
\end{center}
\caption{Evolution of probabilities $P(\nu_{e} \rightarrow \nu_{e})$
that a neutrino emitted as a $\nu_{e}$ at the neutrinosphere remains a
$\nu_{e}$ at some radius. Upper and lower curves correspond to neutrino
energies of 40 MeV and 5 MeV, respectively. \label{fig:conv_prob}}
\end{figure}

\begin{figure}
\begin{center}
\epsfxsize=6.5cm
\epsffile{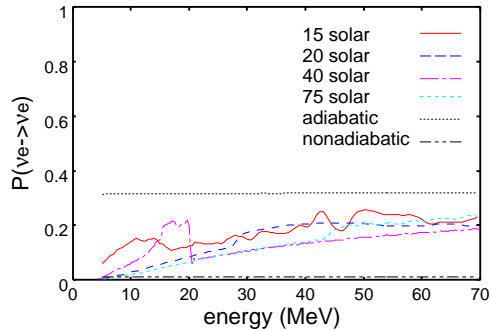}
\end{center}
\caption{Energy dependence of probabilities $P(\nu_{e} \rightarrow
\nu_{e})$ 
that a neutrino emitted as a $\nu_{e}$ at the neutrinosphere remains a
$\nu_{e}$ at the surface of the star. Also shown are the perfectly
adiabatic and non-adiabatic cases for the H-resonance.  
\label{fig:conv_prob_ene}}
\end{figure}

In Fig. \ref{fig:conv_prob}, the evolution of the probability 
$P(\nu_{e} \rightarrow \nu_{e})$ that a neutrino emitted as a $\nu_{e}$
at the neutrinosphere remains a $\nu_{e}$ is shown. The neutrino
energies on the plot are  
5 MeV and 40 MeV. The H-resonance radii and the final probabilities can
be quite 
different for different progenitors and neutrino energies. The
observationally important quantity is the final conversion probability
and this is shown in 
Fig. \ref{fig:conv_prob_ene} as a function of the neutrino energy. The
differences are not so small, about $0.05 - 0.1$ at all energies. But if
$\sin^{2}{2 \theta_{13}}$ is very large or very small, the adiabaticity
parameter is very large or very small, respectively. In this case such
difference of scale height as in Fig. \ref{fig:gradient} will not affect
the neutrino conversion probability. The range of 
$\sin^{2}{2 \theta_{13}}$ is shown to be 
$\sin^{2}{2 \theta_{13}} \lesssim 10^{-5}$ and 
$\sin^{2}{2 \theta_{13}} \gtrsim 10^{-3}$ in \cite{KT4}.

Although the density structure evolves as the shock wave propagates, it
takes about 2 seconds for the shock wave to reach the H-resonance region
\cite{KTshock}. We concentrate our analysis on the early phase when the
shock propagation effect can be neglected. The potential time dependence
of neutrino oscillations due to shock propagation is discussed by several
authors \cite{SchiratoFuller2002,KTshock,LunardiniSmirnov2003,Fogli2003}.

\section{Neutrino Bursts and Neutrino
Oscillations\label{section:oscillation}}

As was stated in the previous section, massive stars experience  
significant mass loss. For current empirical mass loss rates, all
solar-metallicity stars initially more massive than about $35\ M_{\odot}
$ are thought to become hydrogen-free objects of roughly $5 M_{\odot}$
at the end of their thermonuclear evolution. The corresponding upper
limit to the mass of the final iron core is about $2 M_{\odot}$
\cite{WoosleyHegerWeaver2002}.

The mass of the iron core is determined roughly by the Chandrasekhar
mass, which for a zero-temperature, constant $Y_{e}$, Newtonian
structure is \begin{equation} M_{\rm Ch 0} = 5.83 Y^{2}_{e} M_{\odot}.
\end{equation} However, there are numerous corrections, some of which
are large \cite{TimmesWoosleyWeaver1996}. To a first approximation,
the non-zero entropy of the core is important and \begin{equation}
M_{\rm Ch} \sim M_{\rm Ch 0} 
\left[ 1 + \left( \frac{s_{e}}{\pi Y_{e}} \right)^{2} \right],
\end{equation} where \begin{equation} s_{e} = 0.50 \left(
\frac{\rho}{10^{10} {\rm g/cc}} \right)^{-1/3} \left( \frac{Y_{e}}{0.42}
\right)^{2/3} 
\left( \frac{T}{1 {\rm MeV}} \right)
\end{equation}
is the electronic entropy per baryon. More massive stars have higher 
entropy and contain larger iron cores on average. However, 
this general tendency is moderated by the loss and redistribution of
entropy that occurs during the late burning stages. 
Thus, the mass of the iron core as a function of the initial mass will
be somewhat uncertain in that a small change in the initial mass results
in a large difference in the iron core mass. According to
\cite{WoosleyHegerWeaver2002}, the mass of the iron core is $1.2 (1.4) -
1.6 M_{\odot}$ when the initial mass is beween $10 (20) M_{\odot}$ and
$40 M_{\odot}$. This weak dependence 
of the iron core mass on the ZAMS progenitor mass leads to
a somewhat universal neutrino burst. 

\begin{figure}
\begin{center}
\epsfxsize=6.5cm
\epsffile{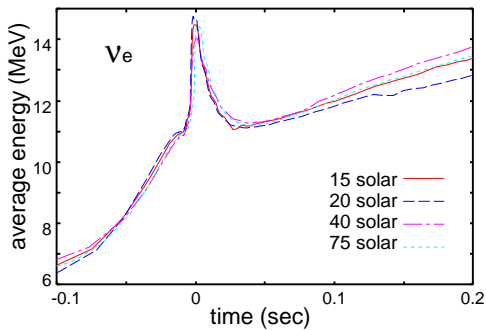}
\end{center}
\caption{Evolution of the average $\nu_{e}$ energy of a neutrino burst
from a progenitor of initial mass $15 M_{\odot}, 20 M_{\odot}, 40
M_{\odot}$ and $75 M_{\odot}$. Time at the bounce is set to zero. 
\label{fig:e-ave}}
\end{figure}

\begin{figure}
\begin{center}
\epsfxsize=6.5cm
\epsffile{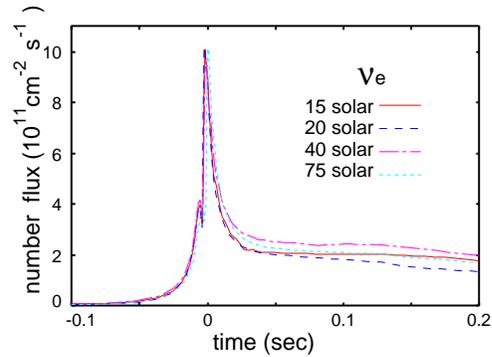}
\end{center}
\caption{Evolution of the number flux at the Earth of $\nu_{e}$
neutrinos due to a neutrino burst from a progenitor  
of initial mass $15 M_{\odot}, 20 M_{\odot}, 40 M_{\odot}$ and $75
M_{\odot}$ at a distance of 10 kiloparsecs (kpc). Time at the bounce is
set to zero. 
\label{fig:e-lum}}
\end{figure}

\begin{figure}
\begin{center}
\epsfxsize=6.5cm
\epsffile{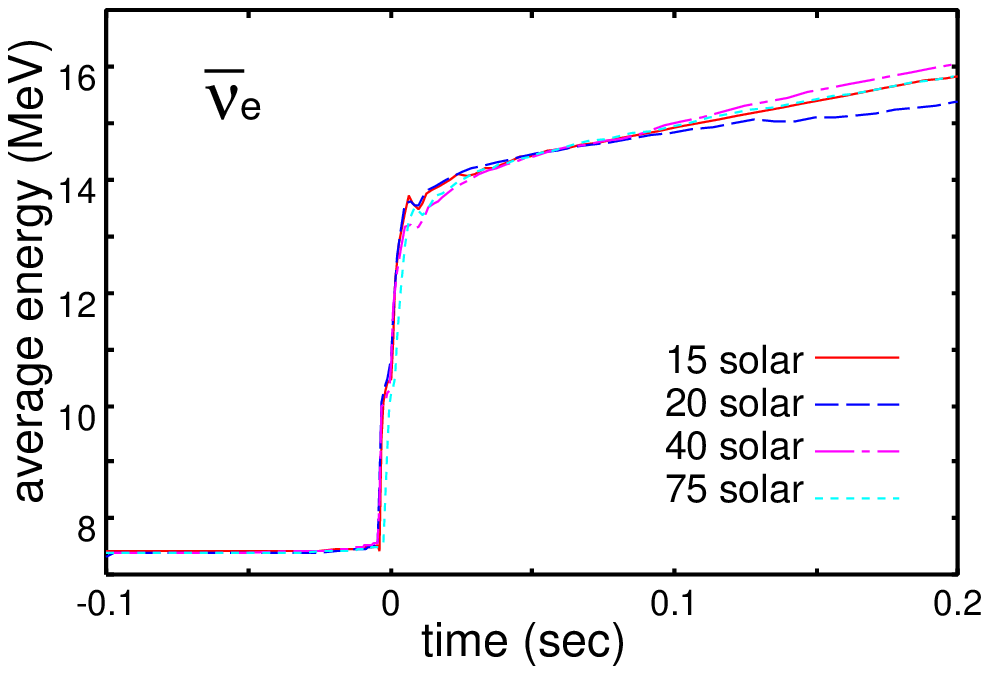}
\end{center}
\caption{Same as Fig. \ref{fig:e-ave}, but for $\bar{\nu}_{e}$.
\label{fig:anti-e-ave}} \end{figure}

\begin{figure}
\begin{center}
\epsfxsize=6.5cm
\epsffile{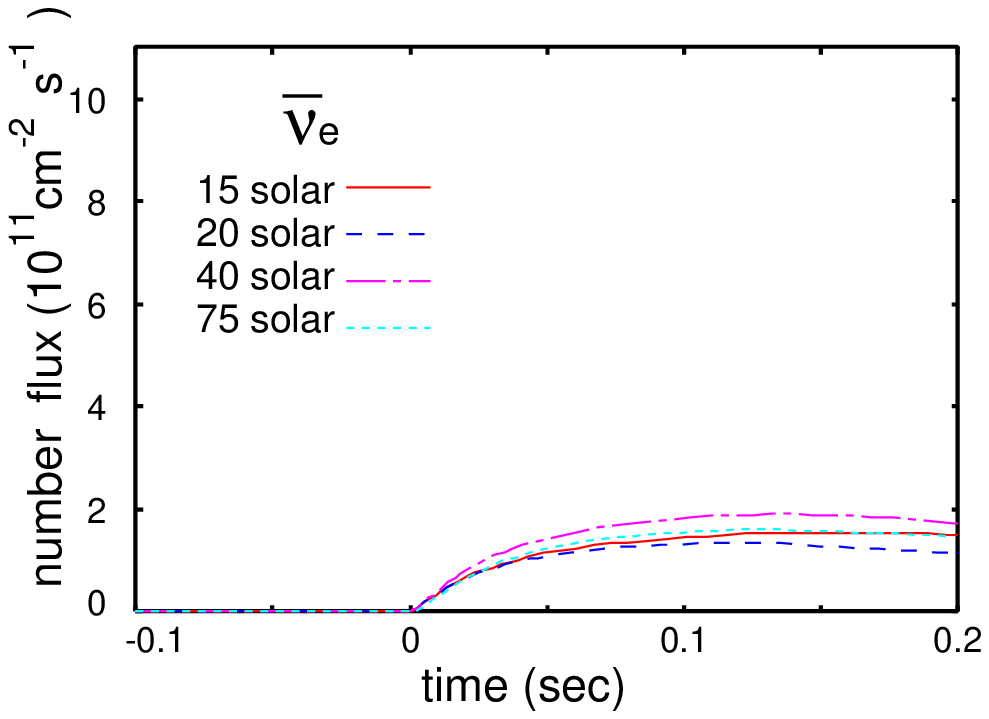}
\end{center}
\caption{Same as Fig. \ref{fig:e-lum}, but for $\bar{\nu}_{e}$.
\label{fig:anti-e-lum}} \end{figure}

\begin{figure}
\begin{center}
\epsfxsize=6.5cm
\epsffile{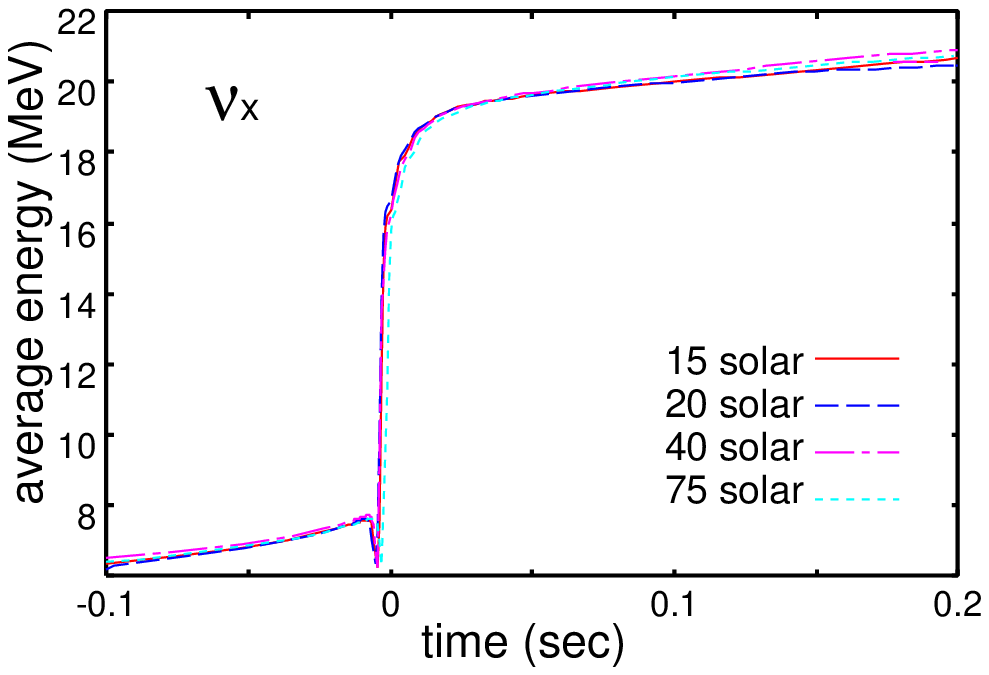}
\end{center}
\caption{Same as Fig. \ref{fig:e-ave}, but for $\nu_{x}$.
\label{fig:x-ave}} \end{figure}

\begin{figure}
\begin{center}
\epsfxsize=6.5cm
\epsffile{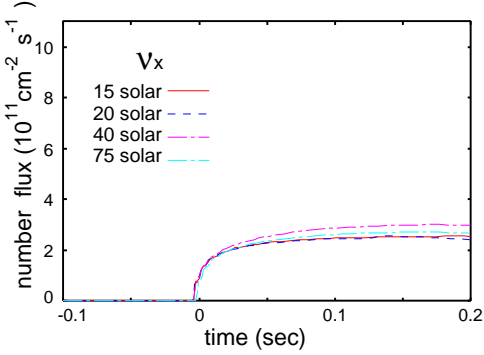}
\end{center}
\caption{Same as Fig. \ref{fig:e-lum}, but for $\nu_{x}$.
\label{fig:x-lum}} \end{figure}

Figs. \ref{fig:e-ave}-\ref{fig:x-lum}
show the evolution of the average neutrino energy and number luminosity
in the early phase up to 200 milliseconds after bounce. The calculation
is based on dynamical models of core-collapse supernovae in one spatial 
dimension, employing a Boltzmann neutrino radiation transport algorithm, 
coupled to Newtonian Lagrangean hydrodynamics and a consistent high-density 
nuclear equation of state. Details of these simulations are described 
in \cite{Burrows}. As can be seen, the major features of the early neutrino 
burst are almost independent of the initial mass.

Combined with the discussion in the previous section, we conclude that
the mantle structure and the features of the neutrino burst depend
little on the initial mass of the progenitor star if $\sin^{2}{2
\theta_{13}} \lesssim 10^{-5}$ or $\sin^{2}{2 \theta_{13}} \gtrsim
10^{-3}$. On one hand, this means that we can not easily obtain 
information about the initial mass from observations of neutrinos 
during the first 200 milliseconds after bounce.  On the other hand, 
this situation is desirable for extracting information about the
neutrino parameters themselves. 

\begin{figure}
\begin{center}
\epsfxsize=6.5cm
\epsffile{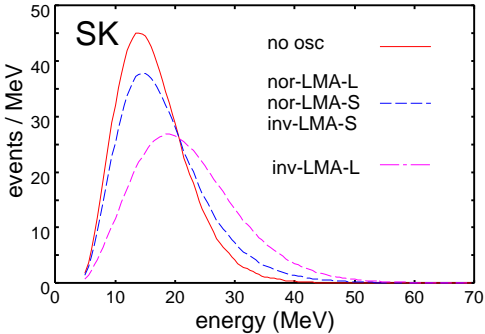}
\end{center}
\caption{Time-integrated event spectra at SuperKamiokande using our 
supernova model with initial mass $20 M_{\odot}$ at a distance of 10
kpc. 
\label{fig:SK_spe}}
\end{figure}

\begin{figure}
\begin{center}
\epsfxsize=6.5cm
\epsffile{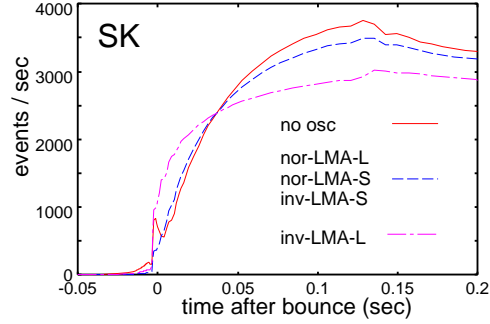}
\end{center}
\caption{Time evolution of the event number in SuperKamiokande at a
distance of 10 kpc using our 
supernova model with an initial mass of $20 M_{\odot}$. 
\label{fig:SK_time}}
\end{figure}

\begin{figure}
\begin{center}
\epsfxsize=6.5cm
\epsffile{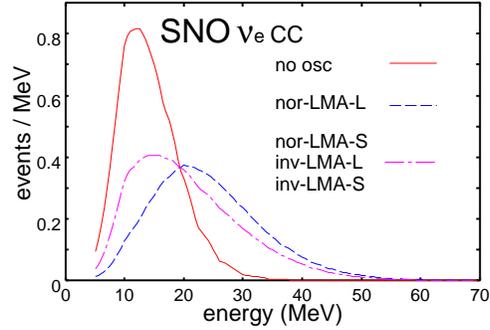}
\end{center}
\caption{Time-integrated event spectra at SNO and a distance of 10 kpc
using our 
supernova model with an initial mass of $20 M_{\odot}$. Only the
charged-current 
events due to the process $\nu_{e} + d \rightarrow p + p + e^{-}$ have
been 
taken into account.
\label{fig:SNO_spe}}
\end{figure}

\begin{figure}
\begin{center}
\epsfxsize=6.5cm
\epsffile{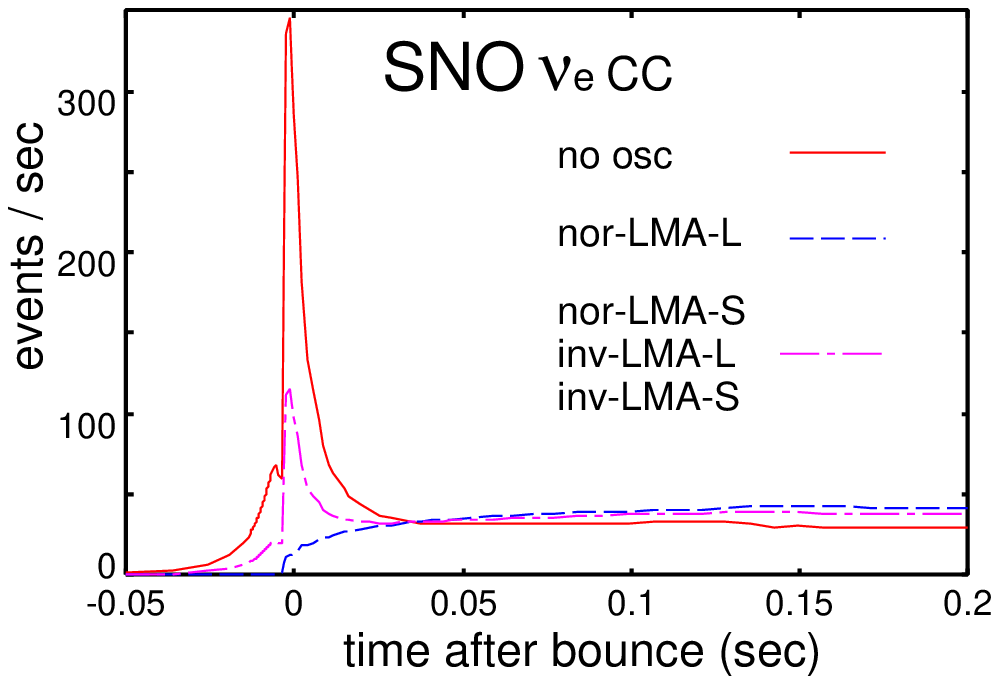}
\end{center}
\caption{Time evolution of the event number at SNO and 10 kpc using our 
supernova model with an initial mass of $20 M_{\odot}$. Only the
charged-current 
events due to $\nu_{e}$ ($\nu_{e} + d \rightarrow p + p + e^{-}$)
absorption have been 
taken into account.
\label{fig:SNO_time}}
\end{figure}

In our previous papers \cite{KT1,KT2,KT3,KT4},
we investigated the possibility of extracting information on neutrino
parameters, especially $\theta_{13}$ and the neutrino mass hierarchy,
from future supernova neutrino observations. We used a numerical
supernova model from the Livermore group \cite{Wilson1986,Totani1998} 
with an initial mass of $15 M_{\odot}$. 
Since the average neutrino energy 
is flavor dependent ($E_{\nu_{e}} < E_{\bar{\nu}_{e}} < E_{\nu_{x}}$),
neutrino oscillations make the spectra of observed $\nu_{e}$ and
$\bar{\nu}_{e}$ neutrinos harder. The extent to which observable neutrinos
become harder depends on the adiabaticity parameters at both the H- and 
L-resonances, which depend on neutrino parameters. 
We considered models with four sets of neutrino parameters:
normal-LMA-L, normal-LMA-S, inverted-LMA-L and inverted-LMA-S. 
Here, normal-LMA-L means that the mass hierarchy is 
normal ($m_{1}^{2} < m_{2}^{2} \ll m_{3}^{2}$) and
$\theta_{13}$ is large ($\sin^{2}{2 \theta_{13}} = 0.043$),
inverted-LMA-S means that the mass hierarchy is inverted ($m_{3}^{2} \ll
m_{1}^{2} < m_{2}^{2}$) and $\theta_{13}$ is small ($\sin^{2}{2
\theta_{13}} = 10^{-6}$), and so on. The other parameters are the same
as in the previous section.

In Figs. \ref{fig:SK_spe}-\ref{fig:SNO_time}, we show for a
20-$M_{\odot}$ progenitor the time-integrated event spectra and the time
evolution of the event number at SuperKamiokande (SK) and SNO. The
events at SNO include only the charged-current events due to the process
$\nu_{e} + d \rightarrow p + p + e^{-}$, while all relevant processes are 
included in the SK event estimates. The detector properties and cross
sections used to calculate these events are described in our previous
paper \cite{KT2}. It should be noted that the neutronization burst
events at SNO are significantly suppressed due to neutrino oscillation. 

In \cite{KT2,KT4}, we used as a measure of neutrino oscillation the 
ratio of high-energy event number to low-energy event number at
SuperKamiokande, whose events 
are dominated by $\bar{\nu}_{e}$, and SNO, which can identify $\nu_{e}$.

High (low) means higher (lower) than 
20 MeV. Our conclusions were that inverted-LMA-L is clearly
distinguishable from the other models although the distinction between
normal-LMA-L and
normal- and inverted-LMA-S is rather difficult. In addition, we
determined that normal-LMA-S and inverted-LMA-S are completely
degenerate \cite{KT4}. One of the uncertainties in our analysis was that
we did not previously take into account the initial progenitor mass
dependence.

\begin{figure}
\begin{center}
\epsfxsize=6.5cm
\epsffile{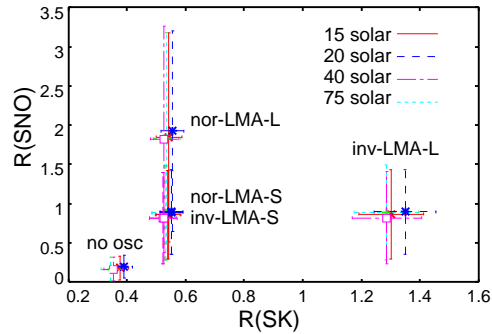}
\end{center}
\caption{Plots of $R_{\rm SK}$ and $R_{\rm SNO}$ for the four neutrino
parameter models mentioned in the text, as well as for the no
oscillation case.  
Results for various initial progenitor masses are given. 
The error bars show statistical errors only.
\label{fig:ratio}}
\end{figure}

We plot in Fig. \ref{fig:ratio} the above-mentioned event-number ratios 
for SK ($R_{\rm SK}$) and SNO ($R_{\rm SNO}$).  Included are results for
the above four neutrino parameter models, as well as for the no oscillation 
case, and various initial 
progenitor masses. Since we use only early-phase events 
($\lesssim 0.2$ sec), the statistical errors are large, whereas in our
previous papers we incorporated the full evolution of the neutrino burst 
based on the simulations of the Livermore group 
\cite{Wilson1986,Totani1998,MayleD,Mayle1987}. 
It can be seen that differences due to initial mass are much smaller
than those due to neutrino parameter and statistical errors. Thus, we
can conclude that although the uncertainty due to the unknown progenitor
mass is not small for intermediate $\theta_{13}$ 
($10^{-5} \lesssim \sin^{2}{2 \theta_{13}} \lesssim 10^{-3}$), we can
nevertheless draw conclusions about whether $\theta_{13}$ is very large
or very small and about the mass hierarchy.

\section{Discussion and Summary\label{section:summary}}

In this paper, we have investigated the initial mass dependence of the
early-phase neutrino signal from supernovae. The early-phase 
analysis has the advantage that it is not affected by the time evolution
of the density structure of the star due to shock propagation and 
is independent of whether the remnant is 
eventually a neutron star or a black hole.
The initial mass does not affect neutrino oscillations
in two senses: the density profile of the mantle, which is important for
neutrino flavor conversion, and the final iron core structure, which
determines the features of the neutrino burst, are almost independent of
the initial mass. This is desirable when we try to extract information
from future supernova-neutrino observations on neutrino parameters.
Although the uncertainty due to the progenitor mass is not small for
intermediate $\theta_{13}$ 
($10^{-5} \lesssim \sin^{2}{2 \theta_{13}} \lesssim 10^{-3}$), we can
nevertheless extract information about whether $\theta_{13}$ is very
large or very small and about the neutrino mass hierarchy.

\section{Acknowledgements}

K.T.'s work is supported by Grant-in-Aid for JSPS Fellows. K.S.'s work
is supported by Grant-in-Aid for Scientific Research (S) No. 14102004
and Grant-in-Aid for Scientific Research on Priority Areas No.
14079202. A.B. acknowledges support from    
the Scientific Discovery through Advanced Computing (SciDAC) program of
the US DOE, grant number DE-FC02-01ER41184.  T.A.T. is supported by NASA
through Hubble Fellowship grant \#HST-HF-01157.01-A awarded by the Space
Telescope Science Institute, which is operated for NASA by the
Association of Universities for Research in Astronomy, Inc., under
contract NAS 5-26555.


\begin{thebibliography}{99}

\bibitem{K2_SN1987a} 
K. Hirata et al., Phys. Rev. Lett. {\bf 58}, 1490 (1987).

\bibitem{IMB_SN1987a} R. M. Bionta et al.,
Phys. Rev. Lett. {\bf 58}, 1494 (1987).

\bibitem{Arafune} J. Arafune and M. Fukugita, 
Phys. Rev. Lett. {\bf 59}, 367 (1987).

\bibitem{Sato}K. Sato and H. Suzuki, 
Phys. Rev. Lett. {\bf 58}, 2722 (1987).

\bibitem{Goldman}I. Goldman et al., 
Phys. Rev. Lett. {\bf 60}, 1789 (1988).

\bibitem{Jegerlehner1996}
B. Jegerlehner, F. Neubig and G. Raffelt, Phys. Rev. D {\bf 54} (1996)
1194.

\bibitem{LunardiniSmirnov}
C. Lunardini and A. Yu. Smirnov, Phys. Rev. D {\bf 63} (2001) 073009.

\bibitem{Minakata2001}
H. Minakata and H. Nunokawa, Phys. Lett. B {\bf 504} 301 (2001).

\bibitem{Fukuda2001}
S. Fukuda et al., Phys. Rev. Lett. {\bf 86} (2001) 5656.

\bibitem{SNO}
SNO Collaboration, Phys. Rev. Lett. {\bf 87} (2001) 071301.

\bibitem{Fukuda1999}
Y. Fukuda et al., Phys. Rev. Lett. 82 (1999) 2644.

\bibitem{KamLAND}
KamLAND Collaboration, K. Eguchi et al., 
Phys. Rev. Lett. {\bf 90} (2003) 021802.

\bibitem{Apollonio1999}
M. Apollonio et al., Phys. Lett. B {\bf 466} (1999) 415.

\bibitem{K2K}
K2K Collaboration, Phys. Rev. Lett. {\bf 90} (2003) 041801.

\bibitem{Dighe2000}
A. S. Dighe and A. Yu. Smirnov, Phys. Rev. D {\bf 62} (2000) 033007.

\bibitem{Fogli2002}
G. L. Fogli, E. Lisi, D. Montanino and A. Palazzo, Phys. Rev. D {\bf 65}
(2002) 073008.

\bibitem{LunardiniSmirnov2003}
C. Lunardini and A. Yu. Smirnov, hep-ph/0302033.

\bibitem{DigheKeilRaffelt1}
A. S. Dighe, M. Th. Keil and G. G. Raffelt, hep-ph/0303210.

\bibitem{DigheKeilRaffelt2}
A. S. Dighe, M. Th. Keil and G. G. Raffelt, hep-ph/0304150.

\bibitem{KT1}
K. Takahashi, M. Watanabe and K. Sato, Phys. Lett. B {\bf 510} (2001)
189.

\bibitem{KT2}
K. Takahashi, M. Watanabe, K. Sato and T. Totani, 
Phys. Rev. D {\bf 64} (2001) 093004.

\bibitem{KT3}
K. Takahashi and K. Sato, Phys. Rev. D {\bf 66} (2002) 033006.

\bibitem{KT4}
K. Takahashi and K. Sato, Prog. Theo. Phys. in press, hep-ph/0205070.

\bibitem{Raffelt2001}
G. G. Raffelt, Astrophys. J. {\bf 561} (2001) 890.

\bibitem{Raffelt2003}
G. G. Raffelt, M. Th. Keil, R. Buras, H. -T. Janka and M. Rampp, Proc.
NOON 03 (10-14 February 2003, Kanazawa, Japan), astro-ph/0303226.

\bibitem{MayleD} 
R. Mayle, Ph. D. Thesis, University of California (1987).

\bibitem{Mayle1987}               
R. Mayle, J. R. Wilson, and D. N. Schramm, Astrophys. J. {\bf 318}
(1987) 288.

\bibitem{WoosleyHegerWeaver2002}
S. E. Woosley, A. Heger and T. A. Weaver, Rev. Mod. Phys. {\bf 74}
(2002) 1015.

\bibitem{Fryer1999}
C. L. Fryer, Astrophys. J. {\bf 522} (1999) 413.

\bibitem{FryerKalogera2001}
C. L. Fryer and V. Kalogera, Astrophys. J. {\bf 554} (2001) 548.

\bibitem{Wilson1986}
J. R. Wilson, R. Mayle, S. Woosley, T. Weaver,
{Ann. NY Acad. Sci. {\bf 470}, 267 (1986)}. 

\bibitem{Totani1998}
T. Totani, K. Sato, H. E. Dalhed and J. R. Wilson, 
Astrophys. J. {\bf 496} (1998) 216.

\bibitem{Burrows}
T. A. Thompson, A. Burrows, and P. A. Pinto, Astrophys. J. {\bf 592},
July 20, 2003 (astro-ph/0211194).

\bibitem{Burrows2}
A. Burrows, Ann. Rev. Nuc. Part.
Sci. {\bf 40} (1990) 181.

\bibitem{Suzuki} 
H. Suzuki: Supernova Neutrino in {\em Physics and Astrophysics of
Neutrino}, 
edited by M. Fukugita and A. Suzuki (Springer-Verlag, Tokyo, 1994).

\bibitem{SCIhomepage}
SciDAC Supernova Science Center homepage: http://www.supersci.org/

\bibitem{Bahcall2003}
J. N. Bahcall, M. C. Gonzalez-Garcia and C. Pena-Garay,
JHEP {\bf 02} (2003) 009.

\bibitem{SchiratoFuller2002}
R. C. Schirato and G. M. Fuller, astro-ph/0205390.  

\bibitem{KTshock}
K. Takahashi, K. Sato, H. E. Dalhed and J. R. Wilson, astro-ph/0212195. 

\bibitem{Fogli2003}
G. L. Fogli, E. Lisi, A. Mirizzi and D. Montanino, hep-ph/0304056.

\bibitem{TimmesWoosleyWeaver1996}
F. X. Timmes, S. E. Woosley and T. A. Weaver, 
Astrophys. J. {\bf 457} (1996) 834.


\end{thebibliography}
\end{document}